%% file: article.tex
\documentclass[12pt]{report}
\include{text/setups/setups}

\usepackage{caption}
\usepackage[english]{babel}
\usepackage{lineno, blindtext}
\usepackage{array}
\usepackage{amsmath}
\usepackage{mdframed}
\usepackage{enumerate}

\usepackage{sectsty}

\chapterfont{\fontsize{24}{15}\selectfont}
\sectionfont{\fontsize{16}{15}\selectfont}
\subsectionfont{\fontsize{14}{15}\selectfont}

\usepackage{background}
\usepackage{lipsum}

\SetBgContents{CLAS12 Note 2017-014}
\SetBgScale{1.35}
\SetBgAngle{0}
\SetBgOpacity{1}
\SetBgColor{black}
\SetBgPosition{current page.south west}
\SetBgHshift{12cm}
\SetBgVshift{20cm}

\begin{document}

 \makeatletter
\newenvironment{sqcases}{%
  \matrix@check\sqcases\env@sqcases
}{%
  \endarray\right.%
}
\def\env@sqcases{%
  \let\@ifnextchar\new@ifnextchar
  \left\lbrack
  \def\arraystretch{1.2}%
  \array{@{}l@{\quad}l@{}}%
}
\makeatother

\newcommand\footnoteref[1]{\protected@xdef\@thefnmark{\ref{#1}}\@footnotemark}

\pdfcompresslevel=9
\pdfobjcompresslevel=9
\include{text/aliases/aliases}

\include{text/title_page/title}

\pagebreak


\include{text/toc/toc}

\include{text/introduction/intro}

\include{text/deut_targ_exp/deut_targ_exp}

\include{text/procedure/procedure}

\include{text/weights/weights}

\include{text/radeff/radeff}

\include{text/concl/concl}


\newpage
\bibliographystyle{ieeetr}
\bibliographystyle{apsrev4-1long}
\bibliography{gen_note}

\end{document}

%% file: text/setups/setups.tex
\usepackage{rotating}
\usepackage[usenames]{color}
\usepackage{overpic}
\usepackage{url}
\usepackage{pst-plot}
\usepackage{amsmath,fullpage,graphicx,caption,float}
\usepackage{amssymb}
\usepackage{array,ragged2e,pst-node,pst-dbicons}
\usepackage{pstricks}
\usepackage{pst-node}
\usepackage{pst-blur}
\usepackage{comment}
\usepackage{color}
\usepackage{subfigure}
\usepackage{lscape}
\usepackage{multirow}
\usepackage{hyperref}
\usepackage{mathtools}
\usepackage{framed}
\usepackage{diagbox}

\usepackage{float}
\restylefloat{table}
\usepackage[margin=25mm]{geometry}

\renewcommand\baselinestretch{1.3}

%% file: text/aliases/aliases.tex
\def\be{\begin{eqnarray}}
 \def\ee{\end{eqnarray}}
 \def\ds{\displaystyle}

\newcommand{\mmm}{\mbox{íÜ÷}}
\newcommand{\reff}[1]{(\ref{#1})}
\newcommand{\ra}{\rangle}
\newcommand{\la}{\langle}
\newcommand{\rf}{\}}
\newcommand{\lf}{\{}
\newcommand{\ket}[1]{ | #1 \rangle }



%% file: text/title_page/title.tex
\noindent\begin{minipage}{\textwidth}
\begin{center}
\thispagestyle{empty}
\vspace{0.5cm}
{ \Large{TWOPEG-D: An Extension of TWOPEG for the Case of\\ a Moving Proton Target}}\\
\vspace{1cm}

{\large Iu. Skorodumina$^{1, a}$, G.V. Fedotov$^{2, 3, b}$,  R.W. Gothe$^{1}$} \\[16pt]

\parbox{.86\textwidth}{\centering\footnotesize\it
$^1$ Department of Physics and Astronomy, University of South Carolina, Columbia, SC 29208.\\
$^2$ Ohio University, Athens, Ohio  45701\\
$^3$ Skobel'tsyn Institute of Nuclear Physics, Moscow State University, Moscow, 119991, Russia\\
E-mail: $^a$ skorodum@jlab.org, $^b$ gleb@jlab.org}\\

\vspace{2cm}
{\bf Abstract}\\[9pt]

\end{center}
{A new TWOPEG-D version of the event generator TWOPEG was developed. This new version simulates the quasi-free process of double-pion electroproduction off the proton that moves in the deuteron target. The underlying idea is the equivalence of the moving proton experiment performed with fixed laboratory beam energy to the proton at rest experiment conducted with effective beam energy different from the laboratory one. This effective beam energy differs event by event and is determined by the boost from the Lab system to the proton rest frame. The Fermi momentum of the target proton is generated according to the Bonn potential. The specific aspects of the deuteron target data analysis are discussed. The plots that illustrate the performance of TWOPEG-D are given. The link to the code is provided. The generator was tested in the analysis of the CLAS data on electron scattering off the deuteron target.}


\end{minipage}

%% file: text/toc/toc.tex
\newpage
\renewcommand{\baselinestretch}{1}\normalsize
\tableofcontents
\setcounter{page}{3}

%% file: text/introduction/intro.tex
\newpage
\chapter{Introduction}
\mbox{}\vspace{-\baselineskip}

During the last decades great efforts have been performed in laboratories all over the world in order to experimentally investigate exclusive reactions of meson electroproduction off the proton. This investigation is typically carried out by detailed analyses of the experimental data with the final goal of extracting various observables. 

By now exclusive reactions off the free proton have been studied in quite detail, and a lot of information about different observables for various exclusive channels have been accumulated~\cite{CLAS_DB}. Meanwhile the exclusive reactions off the deuteron, being less investigated,  start to attract more and more scientific attention, thus causing a strong demand to develop effective tools for their analysis. For this purpose a reliable Monte-Carlo simulation of the process of meson electroproduction off the deuteron target should be elaborated.

This note presents the successful attempt to simulate the quasi-free process of double-pion electroproduction off the proton that moves in the deuteron. The note introduces the TWOPEG-D event generator, which is an extension of the TWOPEG that is the event generator for double-pion electroproduction off the free proton~\cite{twopeg}. In TWOPEG-D the Fermi momentum of the target proton is generated according to the Bonn potential~\cite{Machleidt:1987hj} and then naturally merged into the specific kinematics of double-pion electroproduction.

The basic idea that underlies TWOPEG-D consists in the equivalence of the moving proton experiment performed with fixed laboratory beam energy to the proton at rest experiment conducted  with effective beam energy different from the laboratory one. This effective beam energy differs event by event and is determined by the boost from the Lab system to the proton rest frame and hence depends on the Fermi momentum of the target proton.


TWOPEG-D does not simulate effects of final state interactions (FSI) due to their complexity and not fully understood nature, thus claiming only the ability to imitate the quasi-free process of double-pion electroproduction off moving protons. Beside that, other effects that are  intrinsic to experiments on the bound nucleon (such as the off-shellness of the target nucleon, possible modifications of the reaction amplitudes in the nuclear medium, etc.) are ignored in TWOPEG-D due to their minor significance.

The note is organized in the following way. Section~\ref{sec:data_an_on_mov_p} describes the specific features of a deuteron target experiment, which originate from the fact that the target proton is in motion and cause difficulties during the data analysis. This section also outlines some methods for overcoming these difficulties and demonstrates the essential need for a proper Monte-Carlo simulation of the reaction under investigation. Section~\ref{sect:proc} gives the details of the event generation process and describes the multi-stage procedure of calculating the momenta of the final particles in the Lab frame. The specifity of obtaining the cross section weight is given in Sect.~\ref{sect:weights}, while the details of managing with the simulation of the radiative effects are presented in Sect.~\ref{sect:rad_eff}. The final Section~\ref{sect:concl} contains the link to the repository, where the TWOPEG-D code is located.

It needs to be mentioned that here the reaction is assumed to occur off the proton that moves in the deuteron, but the whole procedure can also be used for any type of the nucleon motion. For instance, if a nucleon  moves inside any nucleus other than deuteron, the Bonn potential should be changed to an appropriate potential of the nucleon-nucleon interaction. Beside that, the procedure can be simply generalized for any exclusive channel.

It also should be emphasized that TWOPEG-D was especially developed to be used in the analyses of data, where the experimental information of the target proton momentum is inaccessible, and one is forced to work under the target-at-rest assumption. If the quality of the experimental data allows to avoid the target-at-rest assumption, it is appropriate to start with the conventional free proton TWOPEG for the Monte-Carlo simulation. 

The user is strongly encouraged to read firstly the note with the detailed description of TWOPEG~\cite{twopeg}, which sketches the kinematics of double-pion electroproduction off the proton, describes the method of event generation with weights, illustrates the quality of the data description, provides details on simulating the radiative effects, etc. This particular note should be treated as an addendum to the report~\cite{twopeg}, since it is fully devoted to the simulation of the effects related to the target motion and no material from the report~\cite{twopeg} is therefore repeated here.

%% file: text/deut_targ_exp/deut_targ_exp.tex
\newpage
\chapter{Specifity of the data analysis off a moving proton}
\label{sec:data_an_on_mov_p}
\mbox{}\vspace{-\baselineskip}

During the deuteron target data analysis one encounters specific issues that are completely external to the free proton data analysis. Those of them that originate solely from the fact of initial proton motion are sketched below.

\section{Fermi smearing of the invariant mass of the initial particles}

For the process of double-pion electroproduction off the proton (as for any other exclusive process) the reaction invariant mass can in general be determined in two ways, i.e. either from the initial particle  four-momenta\footnote[1]{Although the scattered electron is treated as a final particle, here it is classified as ``initial", since it defines the virtual photon, which in turn is attributed to the initial state.} ($W_{i}$) or from the final particle  four-momenta ($W_{f}$) as Eqs.~\eqref{W_fin_1} and~\eqref{W_fin_2} demonstrate\footnote[2]{In electron scattering experiments $W_{i}$ is distorted by the radiative effects, which electrons undergo. The detector resolution also contributes to the difference between experimental values of $W_{i}$ and $W_{f}$.}.

\begin{eqnarray}
W_{i}&= & \sqrt{(P_{p}+P_{\gamma_{v}})^{2}} \label{W_fin_1} \\
W_{f}&= & \sqrt{(P_{\pi^{+}}+P_{\pi^{-}}+P_{p'})^{2}} \label{W_fin_2}
\end{eqnarray}

Here $P_{\pi^{+}}$, $P_{\pi^{-}}$, and $P_{p'}$ are the four-momenta of the final state hadrons, $P_{p}$ is the four-momentum of the initial proton and $P_{\gamma_{v}}=P_{e}-P_{e'}$ the four-momentum of the virtual photon with $P_{e}$ and $P_{e'}$ the four-momenta of the incoming and scattered electrons, respectively.


To determine $W_{f}$, all final hadrons should be registered, while for the calculation of $W_{i}$ it is sufficient to register the scattered electron. 
In the analyses of exclusive reactions, the latter opportunity allows  to use event samples with one unregistered final hadron, whose four-momentum is reconstructed via the missing mass technique. This approach allows to increase the analyzed statistics (sometimes significantly).

The situation complicates for reactions that happen off the proton that moves as in the deuteron. The motion of the target proton is concealed from the observer and usually is not measured. If all particles in the final state are registered, one can restore the information about the momentum of the target proton via the energy-momentum conservation\footnote[3]{In general the target proton momentum can also be reconstructed by measuring the spectator nucleon.}, however if one of the final hadrons is not registered this information turned out to be totally lost. Therefore the value of $W_{i}$ given by Eq.~\eqref{W_fin_1} turns out to be ill-defined, since $P_{p}$ is not known. This brings us to the choice to either demand the registration of all final hadrons to determine $W_{f}$, which reduces the flexibility of the analysis, or to work under a so-called ``target-at-rest assumption", which assumes the initial proton to be at rest. In the last approach the value of $W_{i}$ appears to be smeared.

As a consequence of this smearing, all extracted observables, which depend on the value of $W$, turned out to be convoluted with a function that is determined by the Fermi motion of the initial proton~\cite{Skorodumina:2015rea}. To retrieve the non-smeared observables, a correction that unfolds this effect should be applied. In order to develop this correction, one needs to simulate properly the investigated exclusive process off the moving proton. 

The simulation of $W$-smearing in TWOPEG-D is described in Sect.~\ref{sect:proc}.

\section{Exclusivity cut in the presence of Fermi smearing}

In order to pick out the exclusive reaction, it is a common practice to perform a so-called ``exclusivity cut" as a final step of the event selection. This is a cut on the missing mass, which is calculated via the energy-momentum conservation from the four-momenta of registered particles and reflects the mass spectrum of the unregistered part. For example, for the reaction $ep\rightarrow e'p'\pi^{+}X$, where the scattered electron and final $p$ and $\pi^{+}$ are registered, the missing mass squared of the unregistered part $X$ is determined in the following way,

 \begin{equation}
 M_{X[\pi^{-}]}^{2}=[P_{e} + P_{p}- P_{e'}- P_{p'}-  P_{\pi^{+}}]^{2}.\label{eq:main_top_mm_nosq}
\end{equation}

The investigation of the distribution of the quantity $M_{X[\pi^{-}]}^{2}$ allows to judge the admixture of any types of background as well as the reliability of the entire event selection. A properly chosen position of the exclusivity cut allows at least to suppress the background contribution or even eliminate it completely and to get rid of the non-physical events.

The missing mass is generally subject to the smearing due to the detector resolution. However, if the target proton moves as in the deuteron, the missing mass is also subject to Fermi smearing due to the inevitability to work under the target-at-rest assumption that originates from the incomplete knowledge about the target motion as well as the final hadron state.

If the data analysis includes the estimation of the detector efficiency (for example with the goal to extract a cross section), then the exclusivity cut should also be applied to the reconstructed Monte-Carlo events. In order to calculate the efficiency correctly, the Monte-Carlo distributions should match the experimental ones as well as possible. Fermi smearing of the experimental distributions demands that the simulation should reproduce it. Therefore, the effects of the Fermi motion should be properly included into the Monte-Carlo simulation\footnote[4]{In addition to Fermi smearing, the experimental missing mass distributions are subject to distortions due to FSI effects~\cite{Skorodumina:2015rea}, which can hardly be simulated. This fact increases the importance of the reliable simulation of Fermi smearing for the proper dealing with the FSI contributions during the data analysis. }.

Section~\ref{sect:proc} describes the method that is used in TWOPEG-D for the simulation of the particle four-momenta and gives examples of smeared missing mass distributions.


\section{Transformation to the CMS in the case of moving protons}

For universality purposes the observables are usually extracted in the center-of-mass system (CMS). This implies the transformation of the four-momenta of all particles from the laboratory system (Lab) to the CMS and the subsequent calculation of all kinematical variables from these transformed four-momenta. The description of the kinematical variables for the reaction of double-pion electroproduction off the free proton is given in~\cite{twopeg, Byckling:1971vca,  Fed_an_note:2017}.

The CMS system is uniquely defined as the system, where the initial proton and the photon move towards each other with the $Z_{CMS}$-axis pointing along the photon direction and the net momentum equal to zero. However, the procedure of the Lab to CMS transformation differs depending on the specifity of the reaction's initial state. 

\begin{figure}[!ht]
\begin{center}
\framebox{\includegraphics[width=14cm]{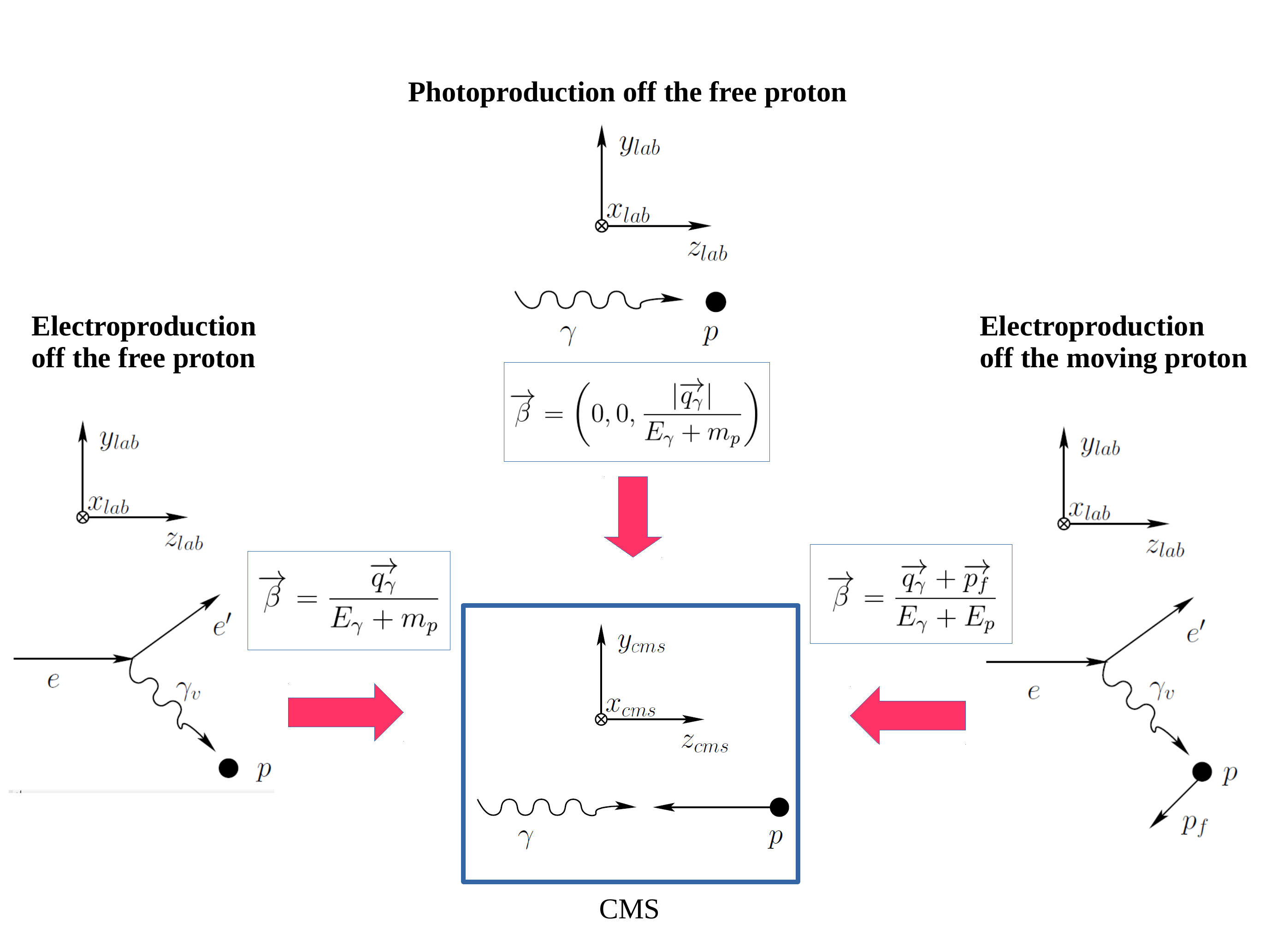}}
\end{center}
\caption{\small  The illustration of three options for the experimental specification of the initial state.}
\label{fig:lab_to_CMS}
\end{figure}

Figure~\ref{fig:lab_to_CMS} illustrates three options\footnote[5]{The fourth option of the reaction off the moving proton induced by the real photons is also possible.} for the experimental specification of the initial state:

\begin{itemize}
\item \textit{The reaction off the free proton induced by the real photons} (upper illustration in Fig.~\ref{fig:lab_to_CMS}). In this case the CMS axis orientation is the same for all reaction events and coincides with that in the Lab system. To transform from the Lab to the CMS, it is sufficient to just perform the boost along the $Z$-axis.
\item \textit{The reaction off the free proton induced by the virtual photons} (left bottom illustration in Fig.~\ref{fig:lab_to_CMS}). In this case the CMS axis orientation is different for each reaction event and is specified by the direction of the scattered electron. To transform from the Lab to the CMS, one needs to perform two rotations to situate the $X$-axis in the electron scattering plane and to align the $Z$-axis with the virtual photon direction. Then the boost along the $Z$-axis can be performed. The analysis report~\cite{Fed_an_note:2017} gives the detailed description of the Lab to CMS transformation for this case.
\item \textit{The reaction off the moving proton induced by the virtual photons} (right bottom illustration in Fig.~\ref{fig:lab_to_CMS}). In this case the CMS axis orientation is again different for each reaction event and is specified by both the scattered electron and the target proton directions. To transform from the Lab to the CMS, one needs to perform the transition to the auxiliary system first, where the proton is at rest, while the incoming electron moves along the $Z$-axis. This transition is determined by the momentum of the target proton. Then the standard procedure described in the previous step can be applied.  
\end{itemize}

Therefore, the need to transform properly from the Lab to the CMS brings us again to the necessity to be aware of the initial proton momentum for each reaction event. If the experiment neither provides the registration of spectator nucleons nor the registration of all final state particles, the correct transformation can not be performed and the extracted observables will lack accuracy. This systematic effect should either be estimated or corrected for. For this purpose the proper simulation of the investigated reaction off the moving proton should be developed.

\section{Ambiguity in the cross section calculation due to its dependence on the beam energy}
\label{sect:ambig}

Electron scattering off the moving proton performed with the beam energy $E_{beam}$ is equivalent to that off the proton at rest conducted with the effective beam energy $\widetilde{E}_{beam}$. This effective beam energy is determined by the boost from the Lab system to the proton rest frame and thus depends on the Fermi momentum of the target proton and differs event by event. Therefore, the experiment off the moving proton with the fixed electron beam energy corresponds to that off the proton at rest performed with the altered beam energy.



The virtual photoproduction cross section $\sigma_{v}$, being decomposed into the combination of the structure functions, has a specific dependence on the beam energy -- the structure functions themselves do not depend on the beam energy, while the dependence is explicitly factorized by the coefficients in front of them\footnote[6]{ For the case of the unpolarized electron beam and the $\pi^{+}\pi^{-}p$ final state this decomposition is given by Eq.~(2.5) of the report~\cite{twopeg}.}. These coefficients incorporate the information about the virtual photon polarization -- they are expressed via the quantities $\varepsilon_{T}$, $\varepsilon_{L}$ or their combinations, where $\varepsilon_{T}$, $\varepsilon_{L}$ are the degrees of transverse and longitudinal polarization of the virtual photon, respectively.

The quantities $\varepsilon_{T}$ and $\varepsilon_{L}$ can be determined according to the following relations\footnote[7]{ $\varepsilon_{T}$ and $\varepsilon_{L}$ given by Eqs.~\eqref{eps_t} and~\eqref{eps_l}  are invariant under the coordinate axis transformation, but not invariant under the Lorentz boost.},

\begin{eqnarray}
\varepsilon_{T}& = &\left( 1 + \frac{Q^{2}\cdot |\overrightarrow{P}_{\gamma_{v}}|^{2}}{2\cdot[\overrightarrow{P}_{e}\times \overrightarrow{P}_{e'} ]^{2}}  \right)^{-1} \textrm{~and}\label{eps_t}\\
\varepsilon_{L}& = &\frac{Q^2}{\nu^2}\varepsilon_{T}\label{eps_l},
\end{eqnarray}
where $\overrightarrow{P}_{\gamma_{v}}$ and $\nu$ are the three-momentum and energy of the virtual photon, respectively, while $\overrightarrow{P}_{e}$ and $\overrightarrow{P}_{e'}$ are the three-momenta of the incoming and scattered electrons, respectively.

Eq.~\eqref{eps_t} gives the general formula for the transverse virtual photon polarization~\cite{Schilling:1973ag}. In the particular case, when the incoming electron moves along the $Z$-axis, this formula is reduced to the well-known expression given by Eq.~(2.3) of the report~\cite{twopeg}, which in turn can be rewritten in the following way to demonstrate the dependence on the beam energy explicitly,

\begin{eqnarray}
\varepsilon_{T} = \frac{1}{1+\frac{2(Q^2+\nu^{2})}{4E_{beam}(E_{beam}-\nu)-Q^2}},\label{eq:dep_on_ebeam}
\end{eqnarray}
where the energy of the virtual photon $\nu$ is fixed for a given $W$ and $Q^2$.

\begin{figure}[!ht]
\begin{center}
\framebox{\includegraphics[width=14cm]{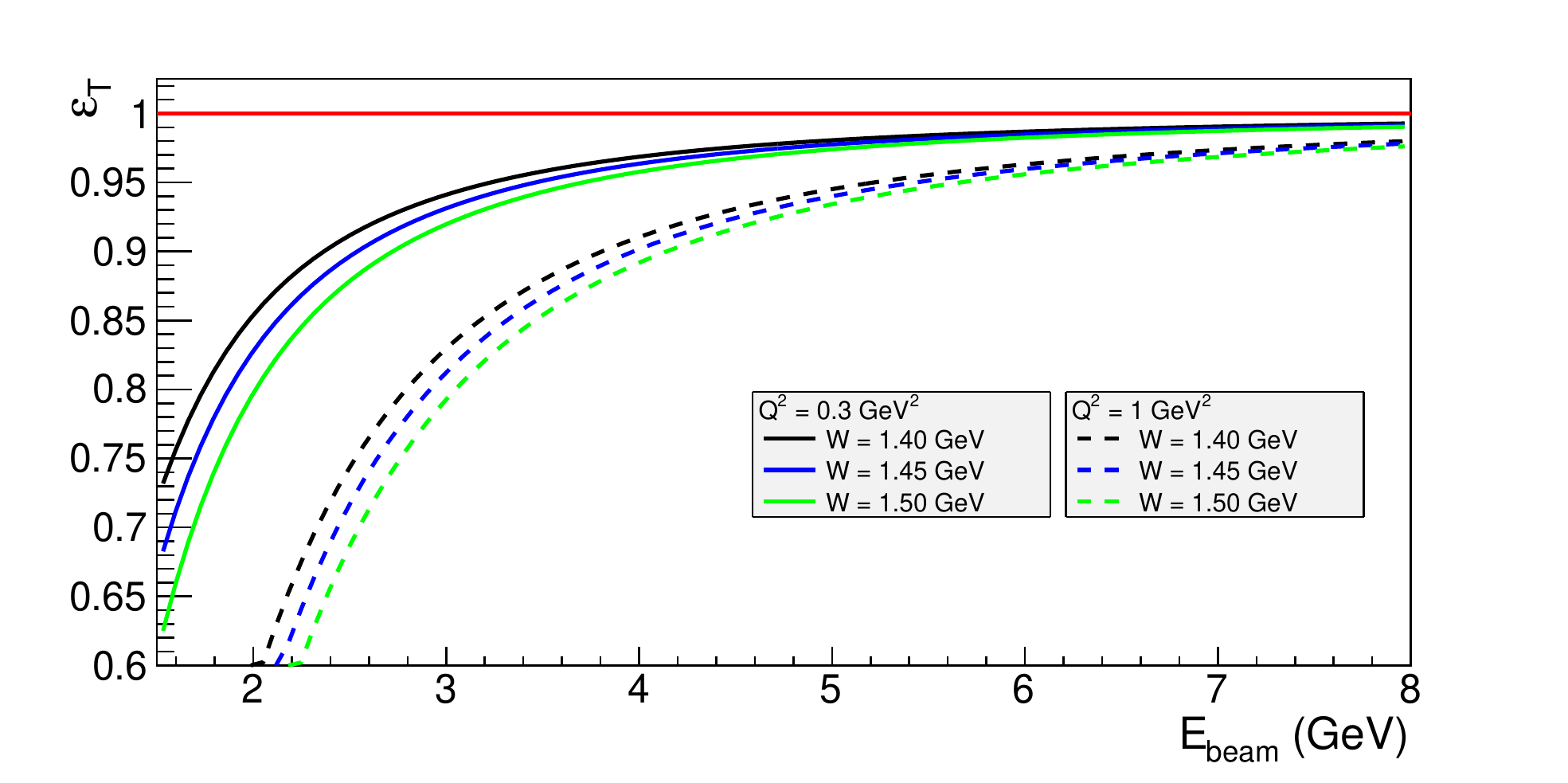}}
\end{center}
\caption{\small  The dependence of $\varepsilon_{T}$ on the beam energy given by Eq.~\eqref{eq:dep_on_ebeam} for the case, when the incoming electron moves along the $Z$-axis in the proton rest frame. The upper bunch of the solid curves corresponds to $Q^{2} = 0.3$~GeV$^2$, while the lower bunch of the dashed curves stands for $Q^{2} = 1$~GeV$^2$. Different colors indicate different fixed values of $W$: 1.4~GeV (black), 1.4~GeV (blue), and 1.5~GeV (green). The red line shows the position of unity.}
\label{fig:eps_t_dep_ebeam}
\end{figure}

Figure~\ref{fig:eps_t_dep_ebeam} illustrates the dependence of $\varepsilon_{T}$ on the beam energy given by Eq.~\eqref{eq:dep_on_ebeam}. The upper bunch of the solid curves corresponds to the fixed $Q^{2} = 0.3$~GeV$^2$, while the lower bunch of the dashed curves stands for  $Q^{2} = 1$~GeV$^2$. Different colors indicate different fixed values of $W$. The higher the beam energy is, the closer the curves are to unity and to each other.

\clearpage

\begin{figure}[!ht]
\begin{center}
\framebox{\includegraphics[width=14cm]{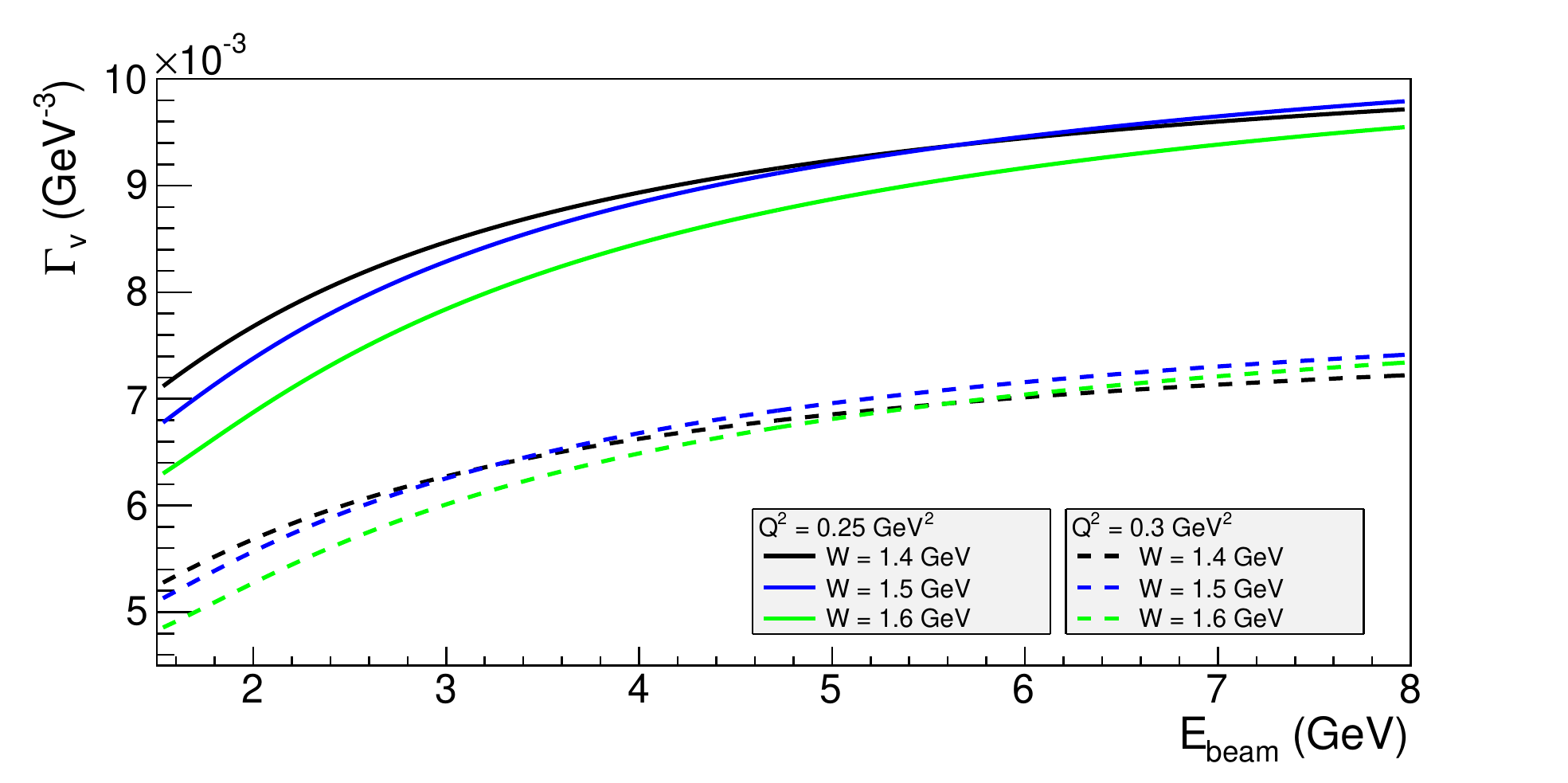}}
\end{center}
\caption{\small  The dependence of $\Gamma_{v}$ on the beam energy for the case, when the incoming electron moves along the $Z$-axis in the proton rest frame. The upper bunch of the solid curves corresponds to $Q^{2} = 0.25$~GeV$^2$, while the lower bunch of the dashed curves stands for $Q^{2} = 0.3$~GeV$^2$. Different colors indicate different fixed values of $W$: 1.4~GeV (black), 1.5~GeV (blue), and 1.6~GeV (green). }
\label{fig:flux_dep_ebeam}
\end{figure}

\clearpage

On top of that, the electroproduction cross section is connected to the virtual photoproduction one via the virtual photon flux $\Gamma_{v}$,  which is also beam energy dependent, as Eq.~(2.2) of the report~\cite{twopeg} demonstrates\footnote[8]{This formula was derived under the assumptions of the incoming electron moving along the Z-axis and the target proton being at rest~\cite{Skorodumina:2016pnb}. }. Figure~\ref{fig:flux_dep_ebeam} illustrates the dependence of the virtual photon flux on the beam energy. The upper bunch of the solid curves again corresponds to the fixed $Q^{2} = 0.25$~GeV$^2$ and the lower bunch of the dashed curves to  $Q^{2} = 0.3$~GeV$^2$. Different colors indicate different fixed values of $W$.

In the proton at rest experiments the conventional practice is to determine $\varepsilon_{T}$, $\varepsilon_{L}$, and $\Gamma_{v}$ in the Lab frame. For the consistency, in the experiments off moving proton these quantities should be defined in the proton rest frame, where the incoming electron has the altered effective beam energy $\widetilde{E}_{beam}$. This circumstance convolutes the extracted cross section with the dependencies of the quantities $\varepsilon_{T}$, $\varepsilon_{L}$, and $\Gamma_{v}$ on the beam energy, hence further complicating the interpretation of the result and its comparison with the cross section of the proton at rest experiment. Although this systematic effect seems not to be significant, it nevertheless should be estimated or corrected for. This can be performed using the proper Monte-Carlo simulation of the reaction under investigation.
 
Section~\ref{sect:proc} describes the calculation of the effective beam energy in TWOPEG-D, while Section~\ref{sect:weights} estimates the influence of the beam energy alteration on the cross section.


\section{Blurring of the $Q^{2}$ versus $W$ distribution boundaries}
\label{sect:blur}

In electron scattering experiments the fixed beam energy imposes kinematical limits on the maximal achievable values of $W$ and $Q^2$. 
The kinematical limitations are usually more strongly restricted by the experimental conditions. One of the experimental restrictions comes from the geometrical limitations of the polar angle of the scattered electron. The boundary of the $Q^{2}$ versus $W$ distribution is then determined by

\begin{eqnarray}
Q^2 &=& \frac{2E_{beam}sin^{2}~\frac{\theta_{e'}}{2}\left (2E_{beam}m_{p}-W^{2}+m_{p}^{2}\right )}{m_{p}+2E_{beam}sin^{2}~\frac{\theta_{e'}}{2}},\label{eq:kin_lim}
\end{eqnarray}
where $m_{p}$ is the proton mass and $\theta_{e'}$ is the polar angle of the scattered electron in the Lab frame.

\begin{figure}[!ht]
\begin{center}
\framebox{\includegraphics[width=14cm]{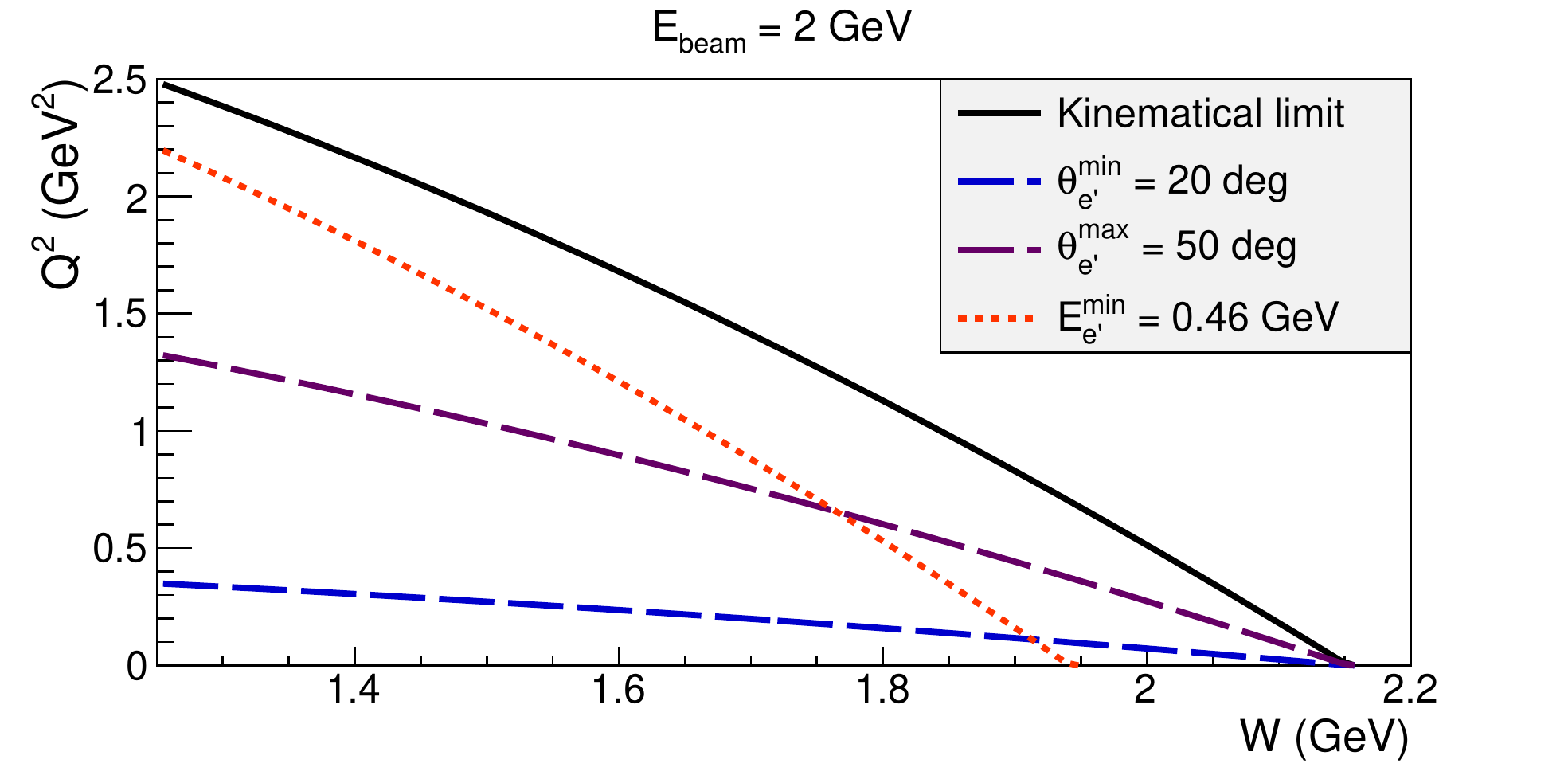}}
\end{center}
\caption{\small  The margins of the $Q^{2}$ versus $W$ distribution for an experiment conducted with 2 GeV beam energy. The solid black curve shows the maximal achievable boundary and is given by Eq.~\eqref{eq:kin_lim} with $\theta_{e'}=180^{\circ}$. The dashed blue and magenta curves stand for the edges due to the limitation of the polar angle of the scattered electron. They are given by Eq.~\eqref{eq:kin_lim} for $\theta_{e'}^{min}=20^{\circ}$ and $\theta_{e'}^{max}=50^{\circ}$, respectively. The dotted red curve shows the edge due to the limitation on the minimal detectable energy of the scattered electron and is given by Eq.~\eqref{eq:kin_lim2} for $E_{e'}^{min} = 0.46$~GeV.}
\label{fig:kin_lim2}
\end{figure}

Figure~\ref{fig:kin_lim2} shows the boundary curves determined by Eq.~\eqref{eq:kin_lim} for $E_{beam} = 2$~GeV and three values of $\theta_{e'}$, i.e. $\theta_{e'}^{min}=20^{\circ}$ (dashed blue), $\theta_{e'}^{max}=50^{\circ}$ (dashed magenta), and $\theta_{e'}=180^{\circ}$ (solid black). The last curve stands for the  maximal achievable limit of the $Q^{2}$ versus $W$ distribution.

Beside that, the experimental coverage can be restricted due to the limitation on the minimal detectable energy of the scattered electron $E_{e'}^{min}$. For this case the boundary curve is given by the following relation,

\begin{eqnarray}
Q^2 &=& m_{p}^{2}+2m_{p}(E_{beam} - E_{e'}^{min}) -W^{2}.\label{eq:kin_lim2}
\end{eqnarray}
 
Figure~\ref{fig:kin_lim2} also shows the boundary curve given by Eq.~\eqref{eq:kin_lim2} for the case $E_{e'}^{min}=0.46$~GeV (dotted red).


The edges of the $Q^{2}$ versus $W$ distribution given by Eqs.~\eqref{eq:kin_lim} and~\eqref{eq:kin_lim2} are beam energy dependent. As written above, the experiment off the moving proton with fixed beam energy is equivalent to that off the proton at rest performed with altered effective beam energy. Therefore, the distribution edges, being sharp and distinct in the proton at rest experiment, become blurred in the experiment off the moving proton.

\begin{figure}[!ht]
\begin{center}
\framebox{\includegraphics[width=14cm]{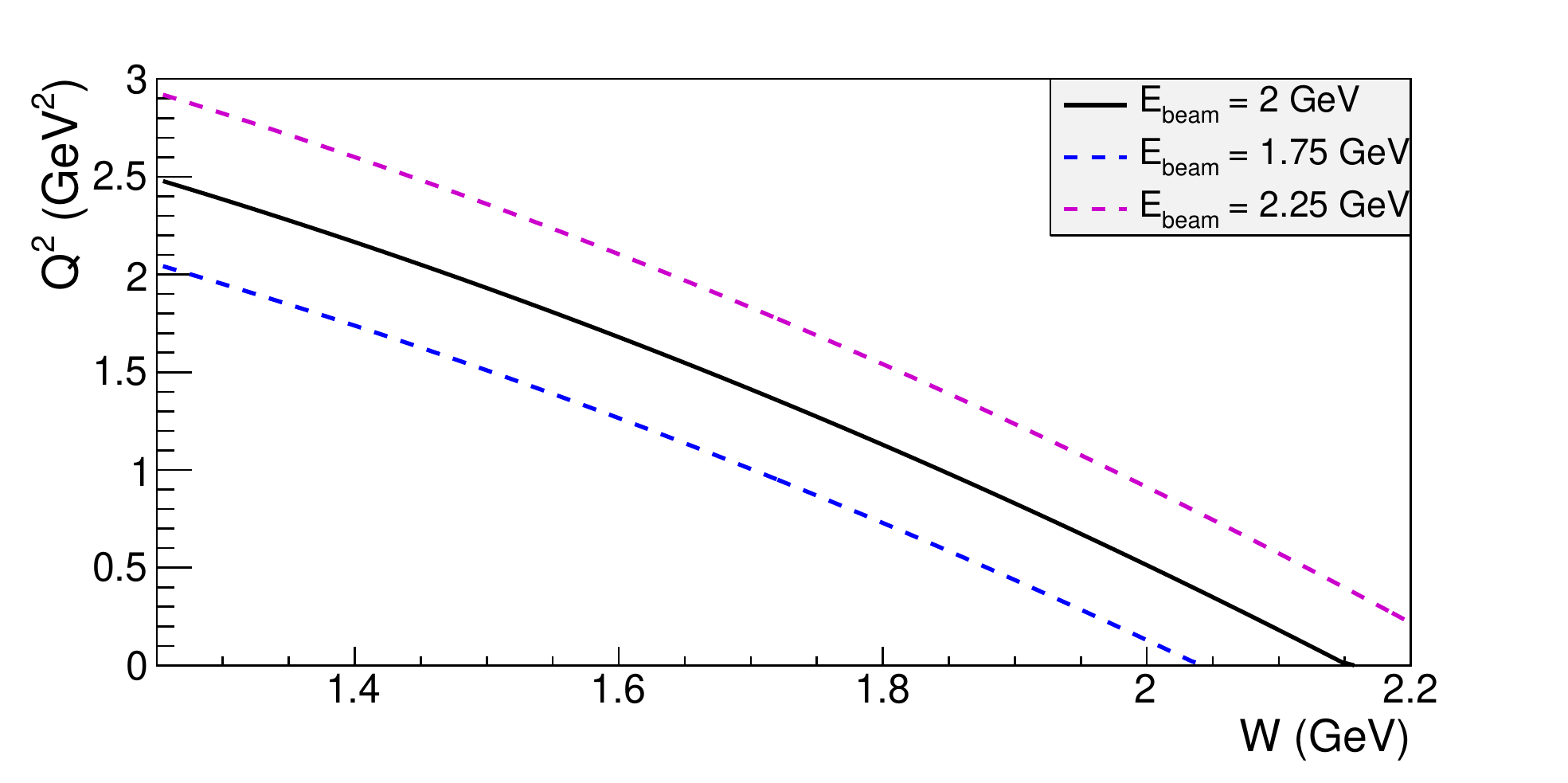}}
\end{center}
\caption{\small  The illustration of blurring of the maximal achievable limit of the $Q^{2}$ versus $W$ distribution. The curves are given by Eq.~\eqref{eq:kin_lim} for the case $\theta_{e'}=180^{\circ}$ and three choices of beam energy. }
\label{fig:kin_lim}
\end{figure}

Let's consider a moving proton experiment conducted with 2 GeV beam energy and assume the deviation of the effective beam energy from this value to be $\pm 250$~MeV. This situation is illustrated in Fig.~\ref{fig:kin_lim}, where the maximal achievable boundaries are shown for three choices of the beam energy: 2 GeV (solid black curve), 1.75 GeV (dashed blue curve), and 2.25 GeV (dashed magenta curve). The region between the two dashed curves shows the scope of the expected blurring. 
The boundaries caused by the experimental restrictions (the dashed and dotted curves in Fig.~\ref{fig:kin_lim2}), being also beam energy dependent, are subject to the analogous blurring.

The event yield in the blurring region suffers from the depletion of events (compared to that for the case of fixed beam energy and sharp disrtribution edge). To estimate this effect, one should know the function that describes the alteration of the effective beam energy. This function is in turn determined by the target proton momentum distribution.
The cross sections extracted in the blurring region need a special correction, otherwise they will suffer from underestimation. This correction requires  either experimental knowledge on initial proton momentum for each reaction event or  the proper Monte Carlo simulation of the blurring effect.

Note that Eqs.~\eqref{eq:kin_lim} and~\eqref{eq:kin_lim2} as well as Figs.~\ref{fig:kin_lim2} and~\ref{fig:kin_lim} assume the value of $W$ to be the true value of the invariant mass of the final hadron system given by Eq.~\eqref{W_fin_2}. Only in this case the boundary blurring takes place. If the smeared value of $W$, calculated under the target-at-rest assumption, is used instead, the distribution edges are not subject to this blurring because the fixed value of the laboratory beam energy is used in calculations.

%% file: text/procedure/procedure.tex
\newpage
\chapter{The event generation procedure}
\label{sect:proc}
\mbox{}\vspace{-\baselineskip}

\section {The generation of the kinematical variables and the Fermi momentum}
\label{sect:gen_var}

For each event the values of all kinematical variables $W$, $Q^2$, $S_{12}$, $S_{23}$, $cos \theta_{h}$, $\phi_{h}$, $\alpha_{h}$ are generated randomly exactly in the same way as it is described in Sect~3.1 of the report~\cite{twopeg}.

\begin{figure}[!ht]
\begin{center}
\framebox{\includegraphics[width=8cm]{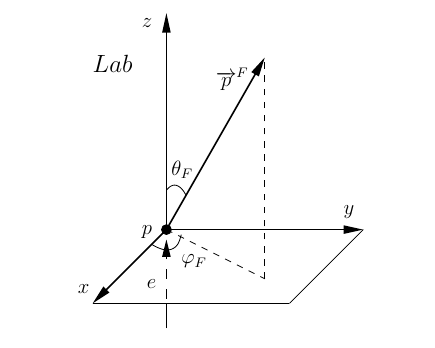}}
\end{center}
\caption{\small The initial conditions of the reaction in the Lab frame. The incoming electron scatters off the proton that moves with the momentum $\protect\overrightarrow{p}^{F}$. }
\label{fig:lab_fermi}
\end{figure}

The simulation of the initial proton motion is performed under the following assumptions. 

\begin{itemize}
\item The Lab frame no longer corresponds to the system, where the target proton is at rest. The target proton moves in the Lab frame with the Fermi momentum as it is shown in Fig.~\ref{fig:lab_fermi}. The axis orientation in the Lab frame is the following: $Z_{lab}$ -- along the beam, $Y_{lab}$ -- up, and $X_{lab}$ -- along $[\vec Y_{lab} \times \vec Z_{lab}]$. 

\item The generated value of $W$ is treated as the smeared one calculated from the initial particle four-momenta according to Eq.~\eqref{W_fin_1} under the target-at-rest assumption (see explanation in Sect.~\ref{sec:data_an_on_mov_p}). Hereinafter this generated value is denoted as $W_{sm}$. The boundaries of the generated $Q^{2}$ versus $W_{sm}$ distribution are set according to Eqs.~\eqref{eq:kin_lim} and~\eqref{eq:kin_lim2} with $E_{beam}$ defined in the Lab frame.

\item The generated value of $Q^{2}$ is treated as the actual $Q^{2}$ value of the reaction. 

\item The four momentum of the incoming electron in the Lab frame is 

\begin{equation}
\begin{aligned}\label{mom_e_ini}
P_{e}^{Lab} = (0, 0, E_{beam}, E_{beam}),
\end{aligned}
\end{equation}
where $E_{beam}$ is the energy of the incoming electron beam that is given as an input parameter.

\item The four-momentum of the scattered electron is defined in the Lab frame exactly in the same way as it is done in the report~\cite{twopeg} (see Eqs.~\eqref{eq:el_in_lab} here as well as Eqs.~(3.2) in the report~\cite{twopeg}). 

\begin{equation}
\begin{split}\label{eq:el_in_lab}
  \nu &= \frac{W_{sm}^2+Q^2-m_{p}^{2}}{2m_{p}}\\
  E_{e'} &= E_{beam}-\nu\\
 \theta_{e'} &= acos\left (1-\frac{Q^2}{2E_{beam}E_{e'}}\right )\\
P_{e'}^{Lab} = (E_{e'}sin \theta_{e'}cos \varphi_{e'}&,E_{e'}sin \theta_{e'}sin \varphi_{e'},E_{e'}cos \theta_{e'},E_{e'}).
\end{split}
\end{equation}

Here $\nu$ is the virtual photon energy in the Lab frame, $m_{p}$ the target proton mass, and $E_{e'}$ and $\theta_{e'}$ the scattered electron energy and polar angle, respectively. $W_{sm}$, $Q^2$, and $\varphi_{e'}$ are the generated reaction invariant mass, the photon virtuality, and the azimuthal angle of the scattered electron, respectively.


The electron defined by Eqs.~\eqref{eq:el_in_lab} imitates the actual scattered electron experimentally registered after the reaction of double-pion electroproduction off the moving proton has happened.

\end{itemize}

The components of the initial proton three-momentum $p_{x}^{F}$, $p_{y}^{F}$, and $p_{z}^{F}$ are generated randomly\footnote[1]{The algorithm of generating the initial proton three-momentum is coded in the subroutine \textit{fermi\_bonn.cxx}.} according to the Bonn potential~\cite{Machleidt:1987hj}. The four-momentum of the initial proton in the Lab frame is then determined by

\begin{equation}
\begin{aligned}\label{mom_p_ini}
P_{p}^{Lab} = (p_{x}^{F}, p_{y}^{F}, p_{z}^{F}, \sqrt{m_{p}^{2}+[p_{x}^{F}]^{2}+[p_{y}^{F}]^{2}+[p_{z}^{F}]^{2}}).
\end{aligned}
\end{equation}

The actual value of the invariant mass of the final hadron system is then determined by\footnote[2]{The determination of $W_{true}$ according to Eq.~\eqref{w_fermi_nonsm} distorts the flatness of the unweighted event distribution of $W_{true}$. This question is addressed in Sect.~\ref{sect:weights}.}

\begin{equation}
\begin{aligned}\label{w_fermi_nonsm}
W_{true}=\sqrt{(P^{Lab}_{p}+P_{\gamma_{v}}^{Lab})^{2}},
\end{aligned}
\end{equation}
where $P^{Lab}_{p}$ is the four-momentum of the moving initial proton defined by Eq.~\eqref{mom_p_ini} and $P_{\gamma_{v}}^{Lab}=P_{e}^{Lab}-P_{e'}^{Lab}$ the four-momentum of the virtual photon with $P_{e}^{Lab}$ and $P_{e'}^{Lab}$ the four-momenta of the incoming and scattered electrons defined by Eqs.~\eqref{mom_e_ini} and~\eqref{eq:el_in_lab}, respectively. 

The components of the initial proton three-momentum are generated under the condition $W_{true}>1.2375$~GeV thus demanding the actual invariant mass of the final hadronic system to be greater than the double-pion production threshold.

\begin{figure}[!ht]
\begin{center}
\framebox{\includegraphics[width=8.8cm]{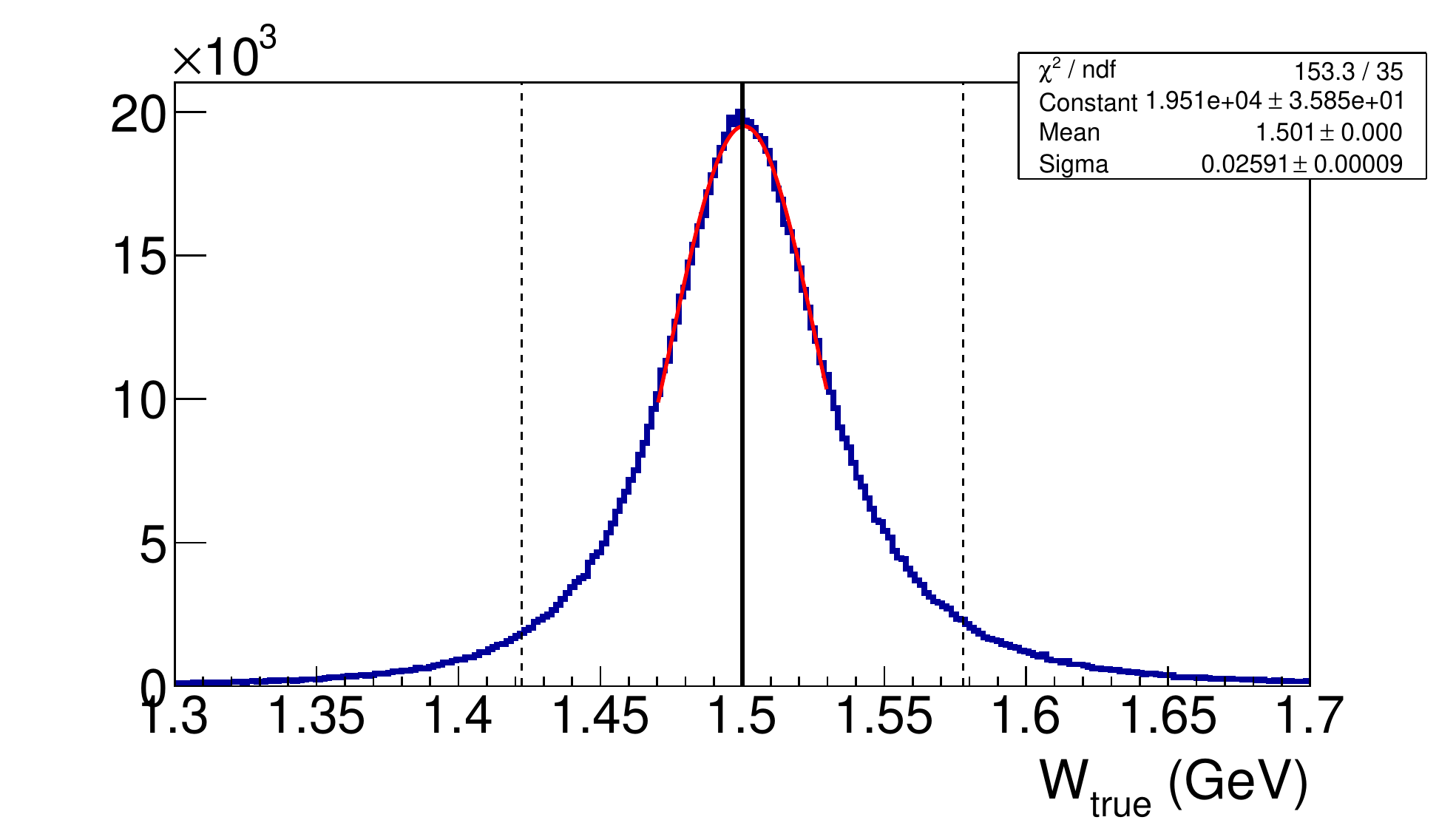}}
\end{center}
\caption{\small The unweighted distribution of $W_{true}$ for the fixed value of $W_{sm} =1.5$~GeV (marked by the solid vertical line). The red curve stands for the Gaussian fit, while the dashed vertical lines mark the values $W_{sm}\pm3\sigma$ to illustrate the distribution's spread. The example is given for $E_{beam} = 2$~GeV and 0.4~GeV$^2$ $< Q^{2}<$ 0.5~GeV$^2$.}
\label{fig:w_smear}
\end{figure}

The scope of Fermi smearing of $W$ is illustrated in Fig.~\ref{fig:w_smear}, which shows the unweighted distribution of $W_{true}$ for the fixed value of $W_{sm} =1.5$~GeV (marked by the solid vertical line). The red curve stands for the Gaussian fit, while the dashed vertical lines mark the values $W_{sm}\pm3\sigma$ to illustrate the distribution's spread. It is seen that the majority of events deviates from the value $W_{sm} =1.5$~GeV within 75~MeV.

\section{Obtaining the particle four-momenta in the Lab frame}

The generated values of the kinematical variables should be used to obtain the four-momenta of all final particles in the Lab frame. For the case of the free proton target the recipe for this is described in Sect.~3.2 of the report~\cite{twopeg} . However, it can not be straightforwardly used for the case of the reaction off the moving proton. Therefore the following multistage method has been developed.

\begin{enumerate}[I.]
\item The four-momenta of the initial particles should be transformed from the Lab frame to the specific system, where the target proton is at rest, while the incoming electron moves along the $Z$-axis. This system hereinafter is denoted as ``quasi-Lab". The initial conditions of the reaction in the quasi-Lab frame imitate those existing in the Lab frame in the case of the free proton experiment. This circumstance determines the name choice ``quasi-Lab" that was assigned to this system\footnote[3]{The transformation of the initial particle four-momenta to the quasi-Lab frame is coded in the subroutine \textit{fermi\_rot.cxx}.}.
\item The procedure described in Sect.~3.2 of the report~\cite{twopeg} should be applied in order to obtain the four-momenta of the final particles in the quasi-Lab frame.
\item The four-momenta of the final particles should be transformed from the quasi-Lab system to the conventional Lab frame\footnote[4]{The transformation of the final particle momenta from the quasi-Lab frame to the Lab system is coded in the subroutine \textit{fermi\_anti\_rot.cxx}.}.
\end{enumerate}

Each step of this method is described below in more details.

\subsection{Obtaining the initial particle four-momenta in the quasi-Lab frame}
\label{sect:transf}

\begin{figure}[!ht]
\begin{center}
\framebox{\includegraphics[width=16.5cm]{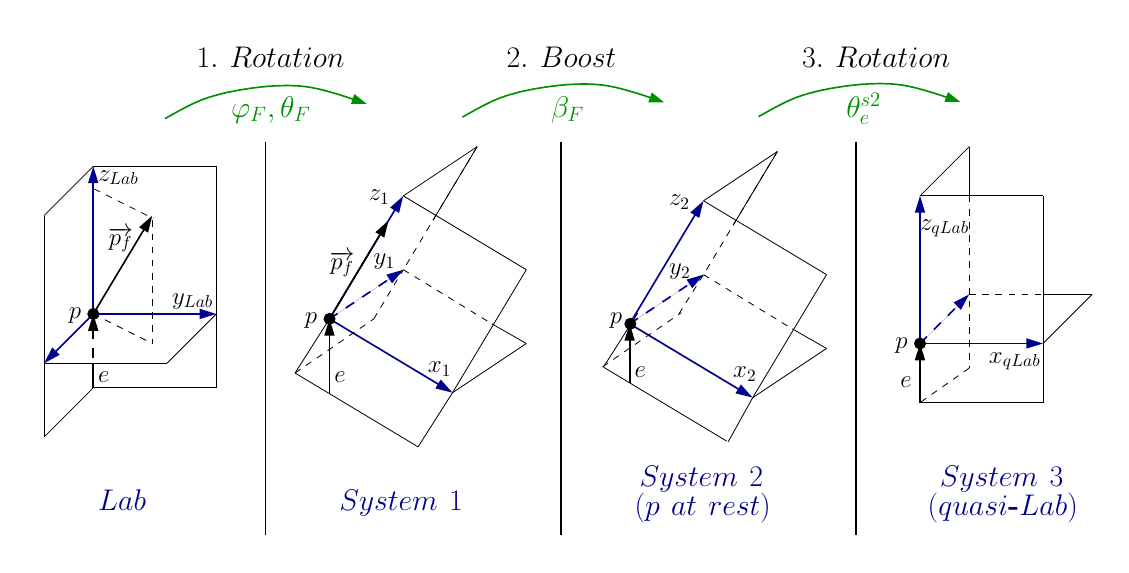}}
\end{center}
\caption{\small Schematical representation of the transformation from the Lab frame to the specific system, where the target proton is at rest, while the incoming electron moves along the $Z$-axis. This system is denoted as ``quasi-Lab". The transformation proceeds via three steps, which are shown by the green arrows.   }
\label{fig:transf_proc}
\end{figure}

The four-momenta of the initial particles defined by Eqs.~\eqref{mom_e_ini} to~\eqref{mom_p_ini} should be transformed from the Lab frame to the quasi-Lab. This transition is performed via three steps, which are schematically shown by the green arrows in Fig.~\ref{fig:transf_proc}. These steps are described below.


\begin{enumerate}
\item The first step is the transformation from the Lab to the auxiliary system, which is denoted in Fig.~\ref{fig:transf_proc} as ``System 1" and represents the frame that has its $Z_{1}$-axis along the target proton momentum. This transformation is performed through a set of rotations of the coordinate axis as described below. 

Firstly the polar $\theta_{F}$ and azimuthal $\varphi_{F}$ angles of the moving initial proton should be calculated in the Lab frame. These angles are marked in Fig.~\ref{fig:lab_fermi} and defined~by~Eq.~\eqref{eq:ang_f}.

\begin{equation}
\begin{aligned}\label{eq:ang_f}
\theta_{F} =&~acos\left (\frac{p_{z}^{F}}{\sqrt{[p_{x}^{F}]^{2}+[p_{y}^{F}]^{2}+[p_{z}^{F}]^{2}}}\right )\\
\widetilde{\varphi}_{F} =&~acos\left (\frac{|p_{x}^{F}|}{\sqrt{[p_{x}^{F}]^{2}+[p_{y}^{F}]^{2}}}\right )\\
\varphi_{F} = &~\begin{sqcases} 
\widetilde{\varphi}_{F},&~~if~~p_{x}^{F}>0~~and~~p_{y}^{F}>0 \\ 
\pi-\widetilde{\varphi}_{F},&~~if~~p_{x}^{F}<0~~and~~p_{y}^{F}>0 \\ 
\widetilde{\varphi}_{F}+\pi,&~~if~~p_{x}^{F}<0~~and~~p_{y}^{F}<0 \\
2\pi-\widetilde{\varphi}_{F},&~~if~~p_{x}^{F}>0~~and~~p_{y}^{F}<0  
\end{sqcases} & \\
\end{aligned}
\end{equation}

Then two subsequent rotations should be made.

The $X_{lab}$-axis is rotated by the angle $\varphi_{F}$ in the $XY$-plane (around the $Z_{lab}$-axis) to force the Fermi momentum to lay in the $XZ$-plane. This rotation translates the axis $Y_{lab}$ to $Y_{1}$ and transforms the four-momentum\footnote[5]{In all derivations the energy is assumed to be the last component of the four-momentum and the four-momentum to be a row vector.} as $P' = P \cdot R_{\varphi_{F}}(\varphi_{F})$ with

\begin{equation}\label{eq:rot_ph_f}
R_{\varphi_{F}}(\varphi_{F}) = \begin{pmatrix}
 cos\varphi_{F}& -sin\varphi_{F} & 0 &0 \\ 
 sin\varphi_{F}& cos\varphi_{F} &  0& 0\\ 
0 & 0 & 1 &0 \\ 
 0&  0&  0&1 
\end{pmatrix}.
\end{equation}

Then one should rotate the $Z_{lab}$-axis by the angle $\theta_{F}$ in the $XZ$-plane in order to translate the axis $Z_{lab}$ to $Z_{1}$ and direct it along the Fermi momentum. This rotation transforms the four-momentum as $P'' = P' \cdot R_{\theta_{F}}(\theta_{F})$ with
\begin{equation}\label{eq:rot_th_f}
R_{\theta_{F}}(\theta_{F})=\begin{pmatrix}
cos\theta_{F} &0  &sin\theta_{F}  &0 \\ 
 0& 1 & 0 &0 \\ 
 -sin\theta_{F} &0  &cos\theta_{F}  & 0\\ 
0 &0  & 0 &1 
\end{pmatrix}.
\end{equation}

As it is sketched in Fig.~\ref{fig:transf_proc}, the incoming electron, being transformed into the ``System 1", turns out to be located in the $XZ$-plane.

\item After that the boost from the ``System 1" to the proton rest frame, which is denoted in Fig.~\ref{fig:transf_proc} as the ``System 2" should be performed. The boost transforms the four-momentum as $P''' = P'' \cdot R_{boost}(\beta)$ with

\begin{equation}\label{eq:boost}
R_{boost}(\beta) = \begin{pmatrix}
1 &0  &0  &0 \\ 
0 &1  &0  &0 \\ 
 0&  0& \gamma  &-\gamma \beta  \\ 
 0&  0& -\gamma \beta  & \gamma 
\end{pmatrix}, \, \, \, \beta =\frac{\sqrt{[p_{x}^{F}]^{2}+[p_{y}^{F}]^{2}+[p_{z}^{F}]^{2}}}{\sqrt{m_{p}^{2}+[p_{x}^{F}]^{2}+[p_{y}^{F}]^{2}+[p_{z}^{F}]^{2}}}, \, \, \,  \textrm{and} \,\,\,   \gamma =\frac{1}{\sqrt{1-\beta ^{2}}},
\end{equation}
where $\beta$ is the magnitude and $Z$-component of the three-vector $\overrightarrow{\beta}=(0,0,\beta)$.

In ``System 2" the incoming electron is still located in the $XZ$-plane.


\item Finally, one should rotate the axis of the proton rest frame (``System 2") to find oneself in the quasi-Lab frame (``System 3"), which has its $Z_{qLab}$-axis along the incoming electron. For that purpose the polar angle of the incoming electron in the ``System 2" should be defined by

\begin{equation}
\begin{aligned}\label{eq:ang_ei_sys2}
\theta_{e}^{s2} =&~acos\left (\frac{p_{z}^{e}}{\sqrt{[p_{x}^{e}]^{2}+[p_{y}^{e}]^{2}+[p_{z}^{e}]^{2}}}\right ),\\
\end{aligned}
\end{equation}

where $p_{x}^{e}$, $p_{y}^{e}$, and $p_{z}^{e}$ are the corresponded components of the incoming electron momentum in the ``System 2".

Then the $Z_{2}$-axis should be rotated with the angle $\theta_{e}^{s2}$ in the $XZ$-plane in order to be translated into $Z_{qLab}$, which is directed along the incoming electron momentum. This rotation transforms the four-momentum as $P'''' = P''' \cdot R_{\theta_{e}^{s2}}(\theta_{e}^{s2})$  with

\begin{equation}\label{eq:rot_th_s2}
R_{\theta_{e}^{s2}}(\theta_{e}^{s2})=\begin{pmatrix}
cos\theta_{e}^{s2} &0  &-sin\theta_{e}^{s2}  &0 \\ 
 0& 1 & 0 &0 \\ 
 sin\theta_{e}^{s2} &0  &cos\theta_{e}^{s2}  & 0\\ 
0 &0  & 0 &1 
\end{pmatrix}.
\end{equation}

\end{enumerate}

After all manipulations the four-momenta of the initial particles are written in the quasi-Lab system in the following way,

\begin{equation}
\begin{aligned}\label{eq:in_part_qlab}
P_{p}^{qLab} &= (0, 0, 0, m_{p}),\\[5pt]
P_{e}^{qLab} &= (0, 0, \widetilde{E}_{beam}^{qL}, \widetilde{E}_{beam}^{qL}) \textrm{,~and} \\[5pt]
P_{e'}^{qLab} &= (E_{e'}^{qL}sin \theta_{e'}^{qL}cos \varphi_{e'}^{qL},E_{e'}^{qL}sin \theta_{e'}^{qL}sin \varphi_{e'}^{qL},E_{e'^{qL}}cos \theta_{e'}^{qL},E_{e'}^{qL}),
\end{aligned}
\end{equation}
where $\widetilde{E}_{beam}^{qL}$ is $Z$-component of the incoming electron momentum in the quasi-Lab frame, while $E_{e'}^{qL}$, $\theta_{e'}^{qL}$, and $\varphi_{e'}^{qL}$ are the energy and spatial angles of the scattered electron in the quasi-Lab frame, respectively.

As it is seen from Eqs.~\eqref{eq:in_part_qlab}, in the quasi-Lab system the target proton is at rest, the incoming electron moves along the $Z_{qLab}$-axis, while the scattered electron has a certain known orientation. Thus, the initial conditions of the reaction in the quasi-Lab system perfectly imitate those existing in the Lab system in the case of the free proton experiment. 

\begin{figure}[!ht]
\begin{center}
\framebox{\includegraphics[width=8.8cm]{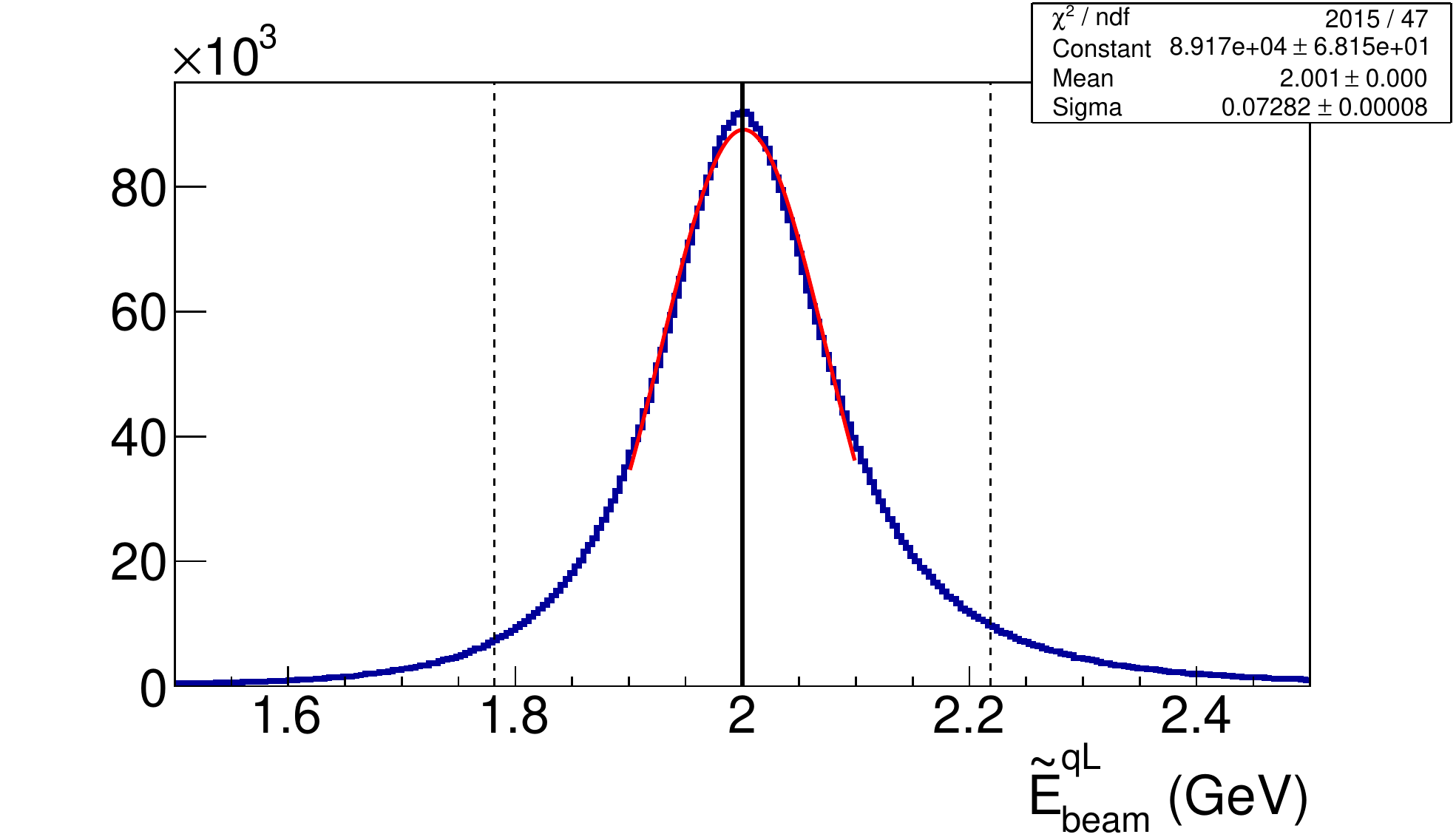}}
\end{center}
\caption{\small The distribution of the effective beam energy $\widetilde{E}_{beam}^{qL}$ defined in the quasi-Lab frame. The solid vertical line shows the value of the beam energy in the Lab frame $E_{beam} = 2$~GeV. The red curve stands for the Gaussian fit, while the dashed vertical lines mark the values $E_{beam}\pm3\sigma$ to illustrate the distribution's spread. The example is given for  1.3~GeV~$<~W_{sm}~<$~1.9~GeV and 0.4~GeV$^2$ $< Q^{2}<$ 0.5~GeV$^2$. }
\label{fig:e_beam_eff}
\end{figure}

$\widetilde{E}_{beam}^{qL}$ in Eqs.~\eqref{eq:in_part_qlab} is a so-called effective beam energy of the incoming electron in the quasi-Lab system, which does not coincide with the usual $E_{beam}$ that is defined in the Lab system and given as an input parameter. This effective beam energy is unique for each event and determined by the generated Fermi momentum.

The distribution of the effective beam energy $\widetilde{E}_{beam}^{qL}$ is shown in Fig.~\ref{fig:e_beam_eff}. The solid vertical line shows the value of the beam energy in the Lab frame $E_{beam} = 2$~GeV. The distribution is almost symmetric with respect to that line\footnote[6]{The minor asymmetry of the distribution comes from the imposed restriction $W_{true}>1.2375$~GeV.}. The red curve stands for the Gaussian fit, while the dashed vertical lines mark the values $E_{beam}\pm3\sigma$ to illustrate the distribution's spread. It is seen that for the majority of events the effective beam energy $\widetilde{E}_{beam}^{qL}$ deviates from the fixed laboratory value within 200~MeV.

As it was discussed in Sect.~\ref{sect:blur}, the alteration of the effective beam energy causes the blurring of the kinematically achievable limits of $W_{true}$ and $Q^2$. TWOPEG-D automatically takes into account this effect, since the calculation of $W_{true}$ according to Eq.~\eqref{w_fermi_nonsm} considers the effective beam energy $\widetilde{E}_{beam}^{qL}$.

Note that the actual invariant mass of the final hadron system $W_{true}$ as well as the photon virtuality $Q^{2}$, being Lorentz invariant, are not subject to any changes during the transformation described above.

\subsection{Obtaining the final hadron four-momenta in the quasi-Lab frame}

The four-momenta of the final hadrons in the quasi-Lab frame are calculated by exactly the same procedure that is described in Sect.~3.2 of the report~\cite{twopeg} for the case of the free proton experiment. 
The procedure should be used as a ``black box" with the following three modifications of its input parameters.

\begin{itemize}
\item One should use the true value of the invariant mass of the final hadron system $W_{true}$ defined by Eq.~\eqref{w_fermi_nonsm} instead of the generated value $W_{sm}$, which is assumed to be smeared.

\item Instead of the true beam energy of the experiment $E_{beam}$, which is defined in the Lab frame, the effective and for each event unique beam energy $\widetilde{E}_{beam}^{qL}$ from Eqs.~\eqref{eq:in_part_qlab} should be used. 

\item Instead of the generated azimuthal angle of the scattered electron $\varphi_{e'}$, which is assumed to be given in the Lab frame, one should use  $\varphi_{e'}^{qL}$ from Eqs.~\eqref{eq:in_part_qlab}, which is defined in the quasi-Lab frame.

\end{itemize}

\subsection{Obtaining the final particle four-momenta in the Lab frame}

Once the four-momenta of the final particles are obtained in the quasi-Lab frame, they should be transformed into the conventional Lab frame. For this purpose they should undergo all transformations shown in Fig.~\ref{fig:transf_proc} in the reverse order. Thus the rule of the four-momentum transformation from the quasi-Lab to Lab is

\begin{equation}\label{eq:qlab_lab_tranf}
P^{Lab}_{i} = P^{qLab}_{i}\cdot R_{\theta_{e}^{s2}}(-\theta_{e}^{s2})\cdot R_{boost}(-\beta) \cdot R_{\theta_{F}}(-\theta_{F}) \cdot R_{\varphi_{F}}(-\varphi_{F}),
\end{equation}
where $P^{Lab}_{i}$ and $P^{qLab}_{i}$ denote the four-momenta of the particle $i$ in the Lab and quasi-Lab frames, respectively. The index $i$ corresponds to $p'$, $\pi^{+}$, $\pi^{-}$, and $e'$. Transformation matrices $R_{\theta_{e}^{s2}}$, $R_{boost}$, $R_{\theta_{F}}$, and $R_{\varphi_{F}}$ are defined by Eqs.~\eqref{eq:rot_th_s2},~\eqref{eq:boost},~\eqref{eq:rot_th_f}, and~\eqref{eq:rot_ph_f}, respectively.

\begin{figure}[!ht]
\begin{center}
\framebox{\includegraphics[width=16cm]{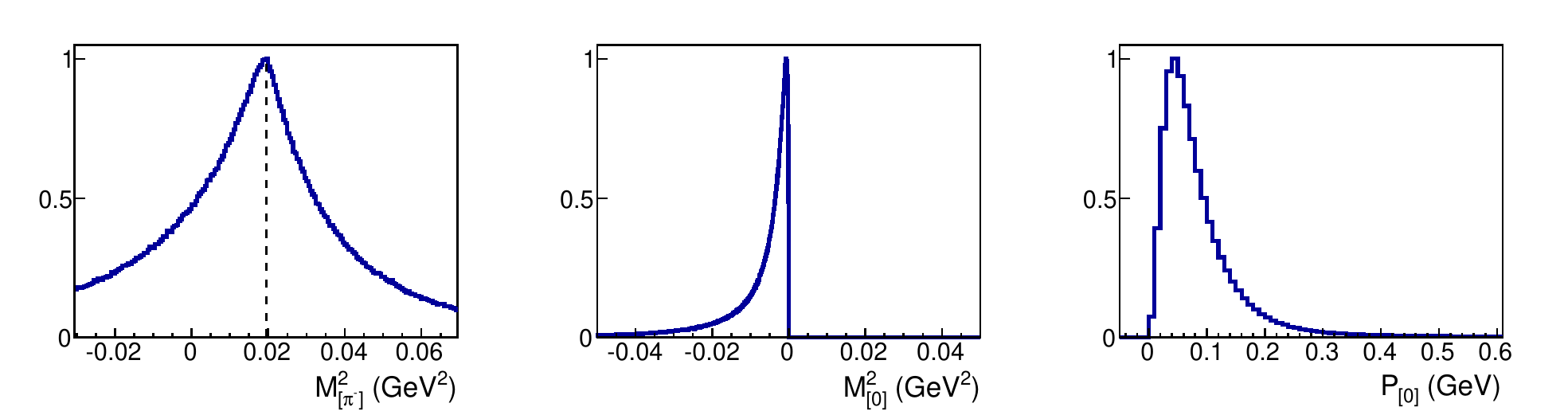}}
\end{center}
\caption{\small The distributions of the quantities $M^{2}_{[\pi^{-}]}$ (left), $M^{2}_{[0]}$ (middle), and $P_{[0]}$ (right), which are defined under the target-at-rest assumption by Eqs.~\eqref{eq:quant_targ_at_rest} and are therefore Fermi smeared. The dashed vertical line in the left plot corresponds to the pion mass squared. The example is given for $E_{beam} = 2$~GeV, 1.3~GeV $< W_{sm} <$ 1.8~GeV, and 0.5~GeV$^2$ $< Q^{2}<$ 0.7~GeV$^2$. }
\label{fig:miss_fermi_smear}
\end{figure}

Figure~\ref{fig:miss_fermi_smear} demonstrates the distributions of the quantities $M^{2}_{[\pi^{-}]}$ (left), $M^{2}_{[0]}$ (middle), and $P_{[0]}$ (right), which are defined in the following way,

\begin{equation}
\begin{aligned}\label{eq:quant_targ_at_rest}
&M^{2}_{[\pi^{-}]} &= &~~[P_{e}^{Lab}&+~&P_{p}&-~&P_{e'}^{Lab}&-~&P_{p'}^{Lab}&-~&P_{\pi^{+}}^{Lab}]^{2}&,\\[5pt]
&M^{2}_{[0]} &= &~~[P_{e}^{Lab}&+~&P_{p}&-~&P_{e'}^{Lab}&-~&P_{p'}^{Lab}&-~&P_{\pi^{+}}^{Lab}&-~&P_{\pi^{-}}^{Lab} ]^{2} \textrm{,~and}\\[5pt]
&~P_{[0]} &= &~~|\overrightarrow{P}_{e}^{Lab}&+~&\overrightarrow{P}_{p}&-~&\overrightarrow{P}_{e'}^{Lab}&-~&\overrightarrow{P}_{p'}^{Lab}&-~&\overrightarrow{P}_{\pi^{+}}^{Lab}&-~&\overrightarrow{P}_{\pi^{-}}^{Lab} |,
\end{aligned}
\end{equation}
where $P_{e}^{Lab}$ and $P_{e'}^{Lab}$ are the four-momenta of the incoming and scattered electrons given by Eqs.~\eqref{mom_e_ini} and~\eqref{eq:el_in_lab}, respectively. $P_{p'}^{Lab}$, $P_{\pi^{+}}^{Lab}$, and $P_{\pi^{-}}^{Lab}$ are the four-momenta of the final hadrons determined by the method described above, while $P_{p}=(0,0,0,m_{p})$ is the four-momentum of the target proton under the target-at-rest assumption. The vectors indicate the corresponding three-momenta.

Equation set~\eqref{eq:quant_targ_at_rest} defines $M^{2}_{[\pi^{-}]}$, $M^{2}_{[0]}$, and $P_{[0]}$ under the target-at-rest assumption in order to imitate the conditions of the real experiment, where the target proton momentum may be not known.  The distributions in Fig.~\ref{fig:miss_fermi_smear}  demonstrate therefore Fermi smearing. The quantity $P_{[0]}$ shown in the right plot, being the missing momentum of the target proton, is distributed according to the Bonn potential~\cite{Machleidt:1987hj}.

%% file: text/weights/weights.tex
\newpage
\chapter{Obtaining the weights}
\mbox{}\vspace{-\baselineskip}
\label{sect:weights}

The weight for each event is determined by exactly the same procedure that is described in Sect.~4 of the report~\cite{twopeg}. The weight factor is calculated according to Eq.~(4.7) from that section with the following three modifications.



\begin{itemize}
\item Instead of the generated value $W_{sm}$, which is assumed to be smeared, the true value $W_{true}$ defined by Eq.~\eqref{w_fermi_nonsm} should be used for picking up the cross section. 

\item To combine the structure functions into the full virtual photoproduction cross section, one should use the values of  $\varepsilon_{T}$ and $\varepsilon_{L}$ calculated in the quasi-Lab frame according to Eqs.~\eqref{eps_t} and~\eqref{eps_l}. See the discussion in Sect.~\ref{sect:ambig}.
\item To obtain the electroproduction cross section from the virtual photoproduction one (mode $F_{flux}=1$), the virtual photon flux $\Gamma_{v}$ should also be calculated in the quasi-Lab system using the effective beam energy $\widetilde{E}_{beam}^{qL}$ introduced by Eqs.~\eqref{eq:in_part_qlab} and the value of $\varepsilon_{T}$ calculated according to Eq.~\eqref{eps_t} in the quasi-Lab frame.
\end{itemize}

\begin{figure}[!ht]
\begin{center}
\framebox{\includegraphics[width=16cm]{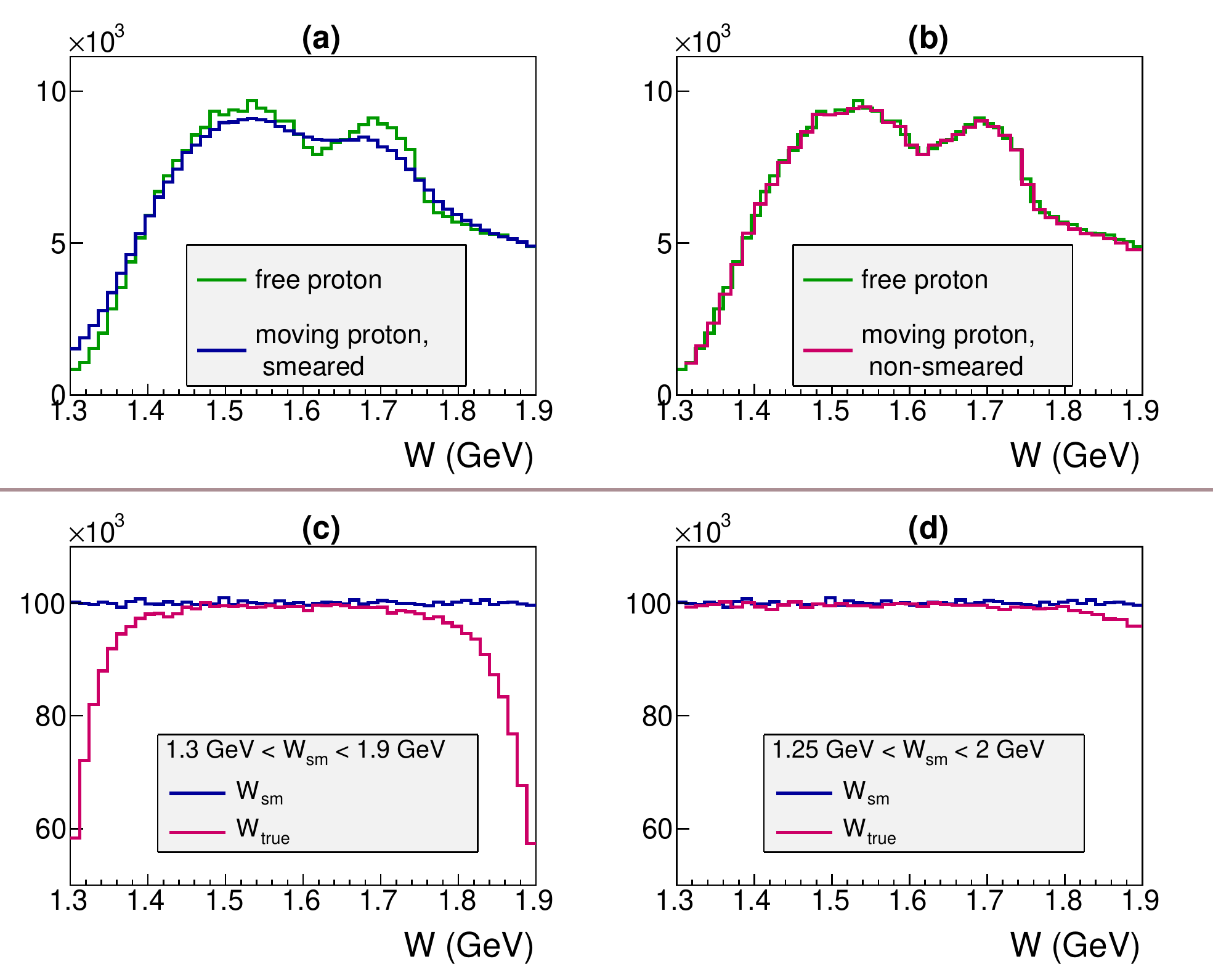}}
\end{center}
\caption{\small {\bf(a)} The comparison of two weighted event distributions, i.e. the $W$ distribution produced by TWOPEG for the case of free proton (green curve) is compared with the $W_{sm}$ distribution produced by TWOPEG-D (blue curve). \\\textbf{(b)} The comparison of two weighted event distributions, i.e. the $W$ distribution produced by TWOPEG for the case of free proton (green curve) is compared with the $W_{true}$ distribution produced by TWOPEG-D (purple curve). See text for more details.\\ \textbf{(c)} The unweighted TWOPEG-D  distributions of $W_{sm}$ generated in a range 1.3~GeV $< W_{sm} <$ 1.9~GeV (blue curve) and the corresponding $W_{true}$ (purple curve).\\\textbf{(d)} The unweighted TWOPEG-D  distributions of $W_{sm}$ generated in a range 1.25~GeV $< W_{sm} <$ 2~GeV (blue curve) and the corresponding $W_{true}$ (purple curve). \\The examples are given for $E_{beam} = 2$~GeV and 0.4~GeV$^{2}$ $<Q^{2}<$ 0.5~GeV$^{2}$.}
\label{fig:weights_w_dep}
\end{figure}

Event distributions that illustrate this procedure are shown in Fig.~\ref{fig:weights_w_dep}. The plot (a) shows the comparison of two weighted event distributions, i.e. the $W$ distribution produced by TWOPEG for the case of free proton (green curve) is compared with the $W_{sm}$ distribution produced by TWOPEG-D (blue curve). The blue curve demonstrates the expected blurring of the resonance structure caused by Fermi smearing. The plot (b) compares the same green curve from free proton TWOPEG with the $W_{true}$ distribution produced by TWOPEG-D (purple curve) and reveals their intrinsic consistency. 

Figure~\ref{fig:weights_w_dep} (b) requires further clarifications. As it is written in Sect.~\ref{sect:gen_var}, TWOPEG-D flatly generates $W_{sm}$, while $W_{true}$ is calculated according to Eq.~\eqref{w_fermi_nonsm} and therefore loses the flatness of generation, being affected by the generation of the Fermi momentum. This is illustrated in Fig.~\ref{fig:weights_w_dep} (c), 
which shows the unweighted TWOPEG-D  distributions of $W_{sm}$ generated in a range 1.3~GeV $< W_{sm} <$ 1.9~GeV (blue curve) and the corresponding $W_{true}$ (purple curve).
While the former distribution is flat, the latter is not: it drops abruptly at the edges and has a plateau in the middle. This behavior is quite justified, since each value of $W_{true}$ can correspond to the sequence of  $W_{sm}$ symmetrically scattered into a certain range (see Fig.~\ref{fig:w_smear}). The Fermi momentum distribution forces most of the $W_{sm}$ values to be located in the vicinity of $W_{true}$ with a deviation of $50$-$100$~MeV, while wider deviations are significantly less probable. Hence, the plateau values of the  $W_{true}$ distribution manage to collect the majority of the corresponded $W_{sm}$ values within whole generated range, while the edge values of $W_{true}$  fail to achieve it. To saturate the edge regions of the $W_{true}$ distribution, the values of $W_{sm}$ should be generated in a wider range, as it is demonstrated in Fig.~\ref{fig:weights_w_dep} (d). To produce this plot, $W_{sm}$ was generated in a range 1.25~GeV $< W_{sm} <$ 2~GeV that leads to the almost full saturation of the $W_{true}$ distribution in a range 1.3~GeV $< W_{true} <$ 1.9~GeV. The comparison of the weighted distributions shown in Fig.~\ref{fig:weights_w_dep} (b) is plotted for the case of saturated unweighted $W_{true}$ distribution shown in Fig.~\ref{fig:weights_w_dep} (d).

The comparison presented in Fig.~\ref{fig:weights_w_dep} (b) demonstrates that the convolution of the cross section with the dependencies of the quantities $\varepsilon_{T}$, $\varepsilon_{L}$, and $\Gamma_{v}$ on the beam energy (see the discussion in Sect.~\ref{sect:ambig}) has an insignificant influence on it. The explanation for that is the following. Due to the fact that the Fermi momentum is directed isotropically, the effective beam energy turned out to be spreaded symmetrically around the actual beam energy with the deviation of $\sim 200$~MeV for the majority of events (see Fig.~\ref{fig:e_beam_eff}). Thus in the limit of high statistics this effect drops out assuming the linear dependence of $\varepsilon_{T}$, $\varepsilon_{L}$, and $\Gamma_{v}$ on the beam energy. The actual dependence of these quantities on the beam energy is demonstrated in Figs.~\ref{fig:eps_t_dep_ebeam} and~\ref{fig:flux_dep_ebeam} and although it is non-linear, in any $\sim 400$~MeV-wide beam energy interval its non-linearity is not pronounced. Therefore, the influence of this effect on the cross section drops out to first order and is negligible in higher orders. 


It needs to be mentioned that TWOPEG-D was especially developed to be used in the analyses of data, where the experimental information of the target proton momentum is inaccessible, and one is forced to work under the target-at-rest assumption. The flat generation of $W_{sm}$ serves this purpose best. If the quality of the experimental data allows to avoid the target-at-rest assumption, one can start with the conventional free proton TWOPEG for the Monte-Carlo simulation. The validity of this proposal is justified by the comparison shown in Fig.~\ref{fig:weights_w_dep} (b).

%% file: text/radeff/radeff.tex
\newpage
\chapter{Managing with radiative effects}
\label{sect:rad_eff}

\mbox{}\vspace{-\baselineskip}

For the simulation of the radiative effects the procedure described in Chapter~7 of the report~\cite{twopeg} was used. However, the task of combining this procedure with the simulation of the target motion is not straightforward. The following two methods were therefore developed and tested.

\section{ Naive method}

In this approach the simulation of the radiative effects was done first, while the simulation of the target motion is performed after that, using the radiated values of $\widetilde{W}$ and $\widetilde{Q^2}$ as well as radiated four-momenta of the incoming and scattered electrons as a starting point.


This method is implemented into the free proton TWOPEG (which also has a complementary moving target mode) and executed under the options $F_{fermi}=1$ and $F_{rad}=1$~or~2.

\section{Advanced method}

In this approach the simulation of the radiative effects is merged with that of the target motion in the following way\footnote[1]{Eqs.~(7.2) to (7.5) are given in the report~\cite{twopeg}.}.

\begin{itemize}

\item The Fermi momentum is generated and the true value of the final hadron system invariant mass $W_{true}$ is calculated according to Eq.~\eqref{w_fermi_nonsm}.

\item The nonradiated cross section in Eqs.~(7.2) to (7.5) are taken for the true value $W_{true}$. To combine transverse and longitudinal structure functions into the full virtual photoproduction cross section and to convert it then to the electroproduction one, the values of $\varepsilon_{T}$, $\varepsilon_{L}$, and $\Gamma_{v}$ were calculated in the quasi-Lab frame.


\item The factor $R_{radsoft}$ in Eq.~(7.2) as well as the integrals given by Eqs.~(7.3) and (7.4) are calculated in the Lab frame. 

\item The maximal allowed energy of the radiated photon ($\omega_{max}^{ini}$ and $\omega_{max}^{fin}$ given by Eqs.~(7.3) and (7.4)) is restricted by the demand to produce a pion pair. This restriction is imposed in the quasi-Lab frame, although the values of $\omega_{max}^{ini}$ and $\omega_{max}^{fin}$ are calculated in the Lab system.

\end{itemize}

This method is implemented into the TWOPEG-D version of the event generator, which always works in the moving target mode ($F_{fermi}=1$).

These two methods turned out to give almost indistinguishable missing mass and momentum distributions and very similar weighted $W$ distributions.
Nevertheless, the second approach is thought to be the preferential one and, therefore, is recommended. 


\begin{figure}[!ht]
\begin{center}
\framebox{\includegraphics[width=16cm]{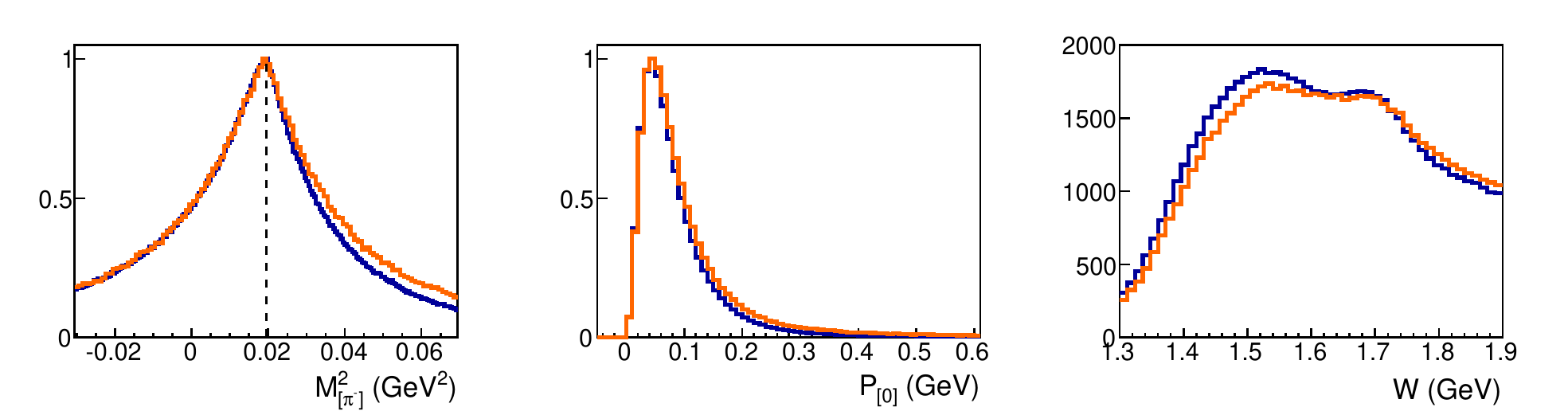}}
\end{center}
\caption{\small  The comparison of the event distributions produced by TWOPEG-D with radiative effects (orange curves) and without (blue curves). The left and middle plots correspond to the quantities $M_{[\pi^{-}]}^{2}$ and $P_{[0]}$, respectively, which were calculated according to Eqs.~\eqref{eq:quant_targ_at_rest}. The right plot shows the comparison of the weighted $W_{sm}$ distributions. The example is given for $E_{beam} = 2$~GeV, 1.3~GeV $< W_{sm} <$ 1.9~GeV, and 0.4~GeV$^2$ $< Q^{2}<$ 0.5~GeV$^2$.  }
\label{fig:w_ferm_rad_norad_comp}
\end{figure}

Figure~\ref{fig:w_ferm_rad_norad_comp} shows the comparison of the event distributions produced by TWOPEG-D with radiative effects (orange curves) and without (blue curves). The left and middle plots correspond to the quantities $M_{[\pi^{-}]}^{2}$ and $P_{[0]}$, respectively, which are given by Eqs.~\eqref{eq:quant_targ_at_rest}. These  quantities were calculated assuming (like in experiment) that $P_{e}^{Lab}$ and $P_{e'}^{Lab}$ are not affected by the radiative effects, while $P_{p'}^{Lab}$, $P_{\pi^{+}}^{Lab}$, and $P_{\pi^{-}}^{Lab}$, on the contrary, take them into account. The right plot shows the comparison of the weighted $W_{sm}$ distributions. 



%% file: text/concl/concl.tex
\chapter{Conclusions and code availability}
\label{sect:concl}

As an extension of TWOPEG~\cite{twopeg} the version TWOPEG-D that simulates the quasi-free process of double-pion electroproduction off a moving proton was developed. 

TWOPEG-D is available as:

\begin{itemize}
\item a separate program TWOPEG-D that works for the case of moving protons only (the mode $F_{fermi}=1$ is fixed). 
\\It can be downloaded at: https://github.com/gleb811/twopeg\_d.git
\item a part of free proton TWOPEG using the mode $F_{fermi}=1$. 
\\It can be downloaded from the same place as specified in the report~\cite{twopeg} \\(i.e., https://github.com/JeffersonLab/Hybrid-Baryons/).
\end{itemize}

With the option $F_{rad}=0$ (without radiative effects) these two editions produce identical results. However, in the mode $F_{rad}=1$ or 2 (with radiative effects) they differ, i.e. the first edition employs the advanced method of merging the radiative effect with the target motion, while the second edition merges them by the naive method, as it is described in more details in Sect.~\ref{sect:rad_eff}.

The specifications of building and running TWOPEG-D are the same as for the free proton TWOPEG. They are described in Sect.~8 of the report~\cite{twopeg}. 

The performance of TWOPEG-D was tested during the analysis of CLAS data on electron scattering off the deuteron target (the part of the ``e1e'' run period)~\cite{Skorodum_wiki_page}, where it has been used for the efficiency evaluation and the corrections due to the radiative effects and Fermi motion of the target proton. For that purpose TWOPEG-D was run in a mode that kept BOS output, which was then passed through the standard CLAS packages GSIM and recsis. In this data analysis TWOPEG-D has proven itself as an effective tool for simulating effects of the target motion for the reaction of double-pion electroproduction off protons.